\documentclass[twoside,twocolumn,english]{revtex4}
\usepackage[T1]{fontenc}
\usepackage[latin9]{inputenc}
\usepackage[a4paper]{geometry}
\makeatletter
\@ifundefined{textcolor}{}
{%
 \definecolor{BLACK}{gray}{0}
 \definecolor{WHITE}{gray}{1}
 \definecolor{RED}{rgb}{1,0,0}
 \definecolor{GREEN}{rgb}{0,1,0}
 \definecolor{BLUE}{rgb}{0,0,1}
 \definecolor{CYAN}{cmyk}{1,0,0,0}
 \definecolor{MAGENTA}{cmyk}{0,1,0,0}
 \definecolor{YELLOW}{cmyk}{0,0,1,0}
}

\usepackage{babel}

\makeatother

\usepackage{babel}
\begin{document}

\title{Comment on ``Generalization of the Kohn-Sham system that can represent
arbitrary one-electron density matrices\textquotedblright \bigskip{}
 }

\author{Mario Piris$^{1,2}$ and Katarzyna Pernal$^{3}$\bigskip{}
 }

\address{$^{1}$Kimika Fakultatea, Euskal Herriko Unibertsitatea (UPV/EHU),
and Donostia International Physics Center (DIPC), 20018 Donostia,
Euskadi, Spain.}

\address{$^{2}$IKERBASQUE, Basque Foundation for Science, 48013 Bilbao, Euskadi,
Spain.}

\address{$^{3}$Institute of Physics, Lodz University of Technology, ul. Wolczanska
219, 90-924 Lodz, Poland.\bigskip{}
 \bigskip{}
 }
\begin{abstract}
Hubertus J. J. van Dam {[}Phys. Rev. A 93, 052512, 2016{]} claims
that the one-particle reduced density matrix (1RDM) of an interacting
system can be represented by means of a single-determinant wavefunction
of fictitious non-interacting particles. van Dam introduced orbitals
within a mean-field framework that produce energy levels similar to
Hartree-Fock (HF) orbital energies, therefore he also claims that
conventional analyses based on Koopmans theorem are possible in 1RDM
functional theory (1RDMFT). In this comment, we demonstrate that both
claims are unfounded. 
\end{abstract}
\maketitle
Inspired by the Kohn-Sham \citep{Kohn1965} practical approach of
the density functional theory (DFT), Hubertus J. J. van Dam has introduced
\citep{VanDam2016} a new kind of generalized orbitals $\left\{ \mathbf{G}_{s}\right\} $
to define an auxiliary Slater determinant in the one-particle reduced
density matrix (1RDM) functional theory (1RDMFT) \citep{Gilbert1975}.
An alternative derivation of such a system of noninteracting particles
was also put forward by Requist and Pankratov \citep{Requist2008}.
In contrast to latter formulation, which resorts to an ensemble of
degenerate Slater determinants, i.e., a mixed state, van Dam presented
a detailed analysis of how a Slater determinant can generate any 1RDM
of an interacting system. van Dam proposes to parametrize the occupation
numbers $\left\{ n_{i}\right\} $ by means of a set of $n_{b}$ orthonormal
vectors $\left\{ \mathbf{C}_{r}\right\} $ (Eq. (5) of \citep{VanDam2016}):
\begin{equation}
\mathbf{C}_{r}\mathbf{C}_{s}={\displaystyle \sum_{i=1}^{n_{b}}C_{ir}^{*}}C_{is}=\delta_{rs}\label{orthoC}
\end{equation}
The columns of the matrix $\mathbf{C}$ are referred to as correlation
functions. In a system with $n_{e}$ electrons ($n_{e}<n_{b}$) of
a given spin, there are $n_{e}$ occupied correlation functions and
all others are unoccupied. Accordingly, the occupation numbers are
given by 
\begin{equation}
{\displaystyle n_{i}={\displaystyle \sum_{r=1}^{n_{e}}C_{ir}^{*}}C_{ir}}={\displaystyle \sum_{r=1}^{n_{e}}}\left|C_{ir}\right|^{2}
\end{equation}
Instead of working with $\left\{ n_{i}\right\} $, imposing the $n_{e}$-representability
conditions on them, van Dam proposes to replace them by the set of
vectors $\left\{ \mathbf{C}_{r}\right\} $. With the above definitions
a new set of $n_{b}$ orthonormal generalized orbitals $\left\{ \mathbf{G}_{s}\right\} $
are defined, where every vector $\mathbf{G}_{s}$ is expanded as (Eq.
(7) of \citep{VanDam2016}) 
\begin{equation}
G_{s}\left(r\right)={\displaystyle \sum_{a,i=1}^{n_{b}}N_{ai}}C_{is}\chi_{a}\left(r\right)\label{Gvector}
\end{equation}
In Eq. (\ref{Gvector}), $\chi_{a}\left(r\right)$ represents the
basis functions and $\left\{ \mathbf{N}_{i}\right\} $ is the orthonormal
set of natural orbitals. Considering the generalized orbitals, van
Dam defines a density matrix $\mathbf{D}$, but unfortunately he made
a mistake in the calculation of its matrix elements, namely, the Eq.
(16) of Ref. \citep{VanDam2016} is not correct. Indeed, the probability
density given a vector $\mathbf{G}_{s}$ according to Eq. (15) of
Ref. \citep{VanDam2016} is 
\begin{equation}
G_{s}\left(r\right)G_{s}^{*}\left(r'\right)=\left({\displaystyle \sum_{a,i=1}^{n_{b}}N_{ai}}C_{is}\chi_{a}\left(r\right)\right)\left({\displaystyle \sum_{b,j=1}^{n_{b}}N_{bj}}C_{js}\chi_{b}\left(r'\right)\right)^{*}\label{GG}
\end{equation}
\[
=\sum_{a,b=1}^{n_{b}}\chi_{a}\left(r\right)D_{ab}^{s}\chi_{b}^{*}\left(r'\right)
\]
where 
\begin{equation}
D_{ab}^{s}=\sum_{i,j=1}^{n_{b}}N_{ai}C_{is}N_{bj}^{*}C_{js}^{*}
\end{equation}

which is different from Eq. (16) of Ref. \citep{VanDam2016}. Note
that only a summation with respect to index $i$ appears in Eq. (16),
instead of the two summations with respect to $i$ and $j$ of equation
(\ref{GG}), respectively. This implies that the results derived subsequently
are also not valid. Particularly, the 1RDM obtained from the Slater
determinant wavefunction, Eq. (17) of Ref. \citep{VanDam2016}, should
read (compare with Eq. (18) of Ref. \citep{VanDam2016}): 
\begin{equation}
\sum_{s=1}^{n_{e}}G_{s}\left(r\right)G_{s}^{*}\left(r'\right)=\sum_{a,b=1}^{n_{b}}\chi_{a}\left(r\right)D_{ab}\chi_{b}^{*}\left(r'\right)
\end{equation}
where 
\begin{equation}
D_{ab}=\sum_{i,j=1}^{n_{b}}N_{ai}\left(\sum_{s=1}^{n_{e}}C_{is}C_{js}^{*}\right)N_{bj}^{*}\label{1RDM}
\end{equation}
contrary to what Eq. (20) of \citep{VanDam2016} shows, namely, 
\begin{equation}
d_{ij}=\sum_{s=1}^{n_{e}}C_{is}C_{js}^{*}\neq\sum_{s=1}^{n_{e}}C_{is}C_{is}^{*}=d_{i}\label{dij}
\end{equation}
The Eqs. (\ref{1RDM}) and (\ref{dij}) imply that the natural orbitals
$\left\{ \mathbf{N}_{i}\right\} $ are not the eigenfunctions of the
1RDM obtained from the determinant wavefunction, as expected. The
genuine 1RDM has a unique set $\left\{ \mathbf{G}_{s}\right\} $ of
eigenfunctions with occupations numbers 0 and 1. The set is unique
up to the rotations among degenerate orbitals. Such a matrix is idempotent,
whereas the 1RDM of an interacting system is never idempotent. The
non-interacting system proposed by van Dam cannot represent the 1-RDM
of any system with fractional occupation numbers.

In order to implement the auxiliary wavefunction, an optimization
scheme was devised based on an \textit{explicit} energy functional
of the 1RDM in both spin channels (Eq. (22) of Ref. \citep{VanDam2016}).
This assumption was addressed at the early stage of the 1RDMFT development
\citep{Donnelly1979,Nguyen-Dang1985}, and recently by Pernal \citep{Pernal2005a}
and Piris and Ugalde \citep{Piris2009a}. Unfortunately, apart from
the special case of the Hartree-Fock (HF) energy expression that may
be viewed as the simplest 1RDM functional, none of the currently known
functionals is explicitly given in terms of the 1RDM, including the
accurate functional describing two-electron closed-shell systems \citep{Goedecker2000},
namely, 
\begin{equation}
\begin{array}{c}
E\left(2e^{-}\right)=2\sum\limits _{p=1}^{\infty}n_{p}\mathcal{H}_{pp}+n_{1}\mathcal{L}_{11}\\
+\sum\limits _{p,q=2}^{\infty}\sqrt{n_{q}n_{p}}\mathcal{L}_{pq}-2{\displaystyle \sum_{p=2}^{\infty}\sqrt{n_{1}n_{p}}\mathcal{L}_{p1}}
\end{array}\label{2e}
\end{equation}

where $n_{p}$ denotes the occupation number in the spatial orbital
$p$, $\mathcal{H}_{pp}$ is the one-electron matrix elements of the
core-Hamiltonian, and $\mathcal{L}_{pq}=\left\langle pp|qq\right\rangle $
is the exchange and time-inversion integral \citep{Piris1999}. The
natural orbital functional (NOF) (\ref{2e}) is obtained from the
exact wavefunction \citep{Shull1959,Kutzelnigg1963} assuming that
all natural occupation amplitudes, with the exception of the first
one, are negative if the first amplitude is chosen to be positive
\citep{Goedecker2000}. It can be readily seen that the electron-electron
interaction energy cannot be explicitly expressed in terms of the
1-RDM due to the different phase factors ($\pm1$) of the occupation
amplitudes. In the case of $N$ electrons, a generalization of this
functional is the extended version of PNOF5 \citep{Piris2013e}, which
can be obtained from a wavefunction of an antisymmetrized product
of strongly orthogonal geminals (APSG) \citep{PernalAPSG}. van Dam
claims that a wide range of energy expressions encompasses the explicit
dependence on the 1RDM, but this is unfounded.

The functionals currently in use are only known in the basis where
the 1RDM is diagonal. This implies that they are not functionals explicitly
dependent on the 1RDM and retain some dependence on the two-particle
reduced density matrix (2RDM). For this reason, it is more appropriate
to speak of a NOF rather than a functional of the 1RDM for an approximate
functional. In this vein, in the NOF theory (NOFT), the natural orbitals
(NOs) are the orbitals that diagonalize the 1RDM corresponding to
an approximate expression of the energy, like those obtained from
an approximate wavefunction.

In NOFT, functionals still depend explicitly on the 2RDM, hence the
energy is not invariant with respect to a unitary transformation of
the orbitals. Consequently, the orbital optimization cannot be reduced
to a pseudo-eigenvalue problem considering the diagonal representation
of the matrix of Lagrange multipliers, associated with the orbital
orthonormality conditions. On this issue, Löwdin already drew attention
in his 1955 landmark paper \citep{Lowdin1955a}. Only if the electron-electron
interaction energy is an explicit functional of the 1RDM, the functional
derivative present in Eq. (29) of Ref. \citep{VanDam2016} may be
directly calculated \citep{Donnelly1979}. In the case of an implicit
functional, the proper procedure to obtain the derivative was proposed
by Pernal \citep{Pernal2005a}, which is based on using the chain
rule and first-order perturbation theory applied to the eigenequation
of the 1RDM. The van Dam's assertion that the secular equation to
determine the NOs (Eq. (30) of Ref. {[}2{]}) is the same as the KS
equation is not applicable in NOFT. Even for the yet unknown exact
ground-state functional of the 1RDM, the secular equation to determine
the NOs bears solely a striking formal resemblance to the HF equations
\citep{Donnelly1979}.

The exact functional of closed-shell two-electron systems given by
the Eq. (\ref{2e}), as well as PNOF5, are approximate but are strictly
N-representable functionals. This points to another issue, that of
the N-representability problem of the functional \citep{Herbert2003a,Piris2010a,Ludena2013}.
The latter refers to the conditions that guarantee the one-to-one
correspondence between $E[\Psi]\equiv E[\mathrm{2RDM}]$ and $E[\mathrm{1RDM}]$.
Several proposals have appeared in the literature \citep{Piris2007,Pernal2016},
in which the 2RDM is expressed in terms of the 1RDM by means of a
reconstruction functional. Accordingly, these reconstructions must
comply with the known conditions for the N-representability of the
2RDM. It has been generally assumed that there is no N-representability
problem of the functional, as it was believed that only N-representable
conditions on the 1-RDM were necessary. The ensemble N-representability
constraints for acceptable 1RDMs are easy to implement, but are insufficient
to guarantee that the reconstructed 2RDM is N-representable, and thereby
the functional either.

The first explicit approximate relation between 2RDM and 1RDM containing
one free parameter was that proposed by Müller in 1984 \citep{Muller1984},
generalized later by Sharma et al. \citep{Sharma2008} as the power
functional, and applied in work \citep{VanDam2016} to Be and LiH.
While it is true that the Müller functional has a simple dependence
upon the 1RDM, it has serious shortcomings \citep{Goedecker2000}.
The matter is that this simple JK-only functional avoids the phase
dilemma discussed above, that stems from the fact that the construction
of the functional requires a choice over a large number of possible
combinations signs in the electron-electron interaction between NOs.
In doing so, the functional seems to depend properly on the 1RDM,
but violates the N-representability conditions for the 2RDM. Several
reconstruction functionals for NOFT, including Müller's, were investigated
by Herbert and Harriman \citep{Herbert2003a}, and illustrative calculations
were precisely made for Be and LiH. They documented extensive N-representability
violations for proposed reconstruction functionals. Although we can
obtain quite reasonable results, these do not guarantee that calculations
made using the power functional are accurate.

Finally, the explicit dependence on the 2RDM in NOFT leads to two
other important consequences. Firstly, the correct procedure in NOFT
for describing the electron detachment in terms of one-electron quantities
is the use of the extended Koopmans' theorem, which provides the connection
between the 1RDM and 2RDM with ionization potentials \citep{Pernal2005,Leiva2006,Piris2012}.
Secondly, the NOFT provides two complementary representations of the
one-electron picture, namely, the NO representation and the canonical
orbital (CO) representation \citep{Piris2013}. The former arises
directly from the optimization process solving the corresponding Euler
equations, whereas the latter is attained from the diagonalization
of the matrix of Lagrange multipliers obtained in the natural orbital
representation. The 1-RDM is diagonal in the NO representation but
not the Lagrangian, which is only a Hermitian matrix. Conversely,
in the CO representation, the Lagrangian is diagonal but not the 1RDM.
It has been shown \citep{Piris2013} by means of the extended Koopmans'
theorem that the one-particle energies associated to the COs can yield
the ionization potentials when the 1RDM remains close to the diagonal
form.

To summarize, the procedure proposed by van Dam for obtaining a noninteracting
system that can represent the one-electron density matrix of any system,
as well as the conventional analysis based on Koopmans' theorem,
are not valid.

Financial support comes from the Spanish MINECO/FEDER Project No.
CTQ2015-67608-P. 

\expandafter\ifx\csname natexlab\endcsname\relax\global\long\def\natexlab#1{#1}
\fi \expandafter\ifx\csname bibnamefont\endcsname\relax \global\long\def\bibnamefont#1{#1}
\fi \expandafter\ifx\csname bibfnamefont\endcsname\relax \global\long\def\bibfnamefont#1{#1}
\fi \expandafter\ifx\csname citenamefont\endcsname\relax \global\long\def\citenamefont#1{#1}
\fi \expandafter\ifx\csname url\endcsname\relax \global\long\def\url#1{\texttt{#1}}
\fi \expandafter\ifx\csname urlprefix\endcsname\relax\global\long\def\urlprefix{URL }
\fi \providecommand{\bibinfo}[2]{#2} \providecommand{\eprint}[2][]{\url{#2}}

\end{document}